\documentclass[sigconf]{acmart}
\pdfoutput=1
\usepackage[utf8]{inputenc}
\usepackage[T1]{fontenc}

\usepackage{algorithm}
\usepackage{algorithmic}
\usepackage{amsmath}
\usepackage{booktabs}
\usepackage{braket}
\usepackage{multirow}
\usepackage{pgfplots}
\usepackage{subfig}
\usepackage{tabularx}
\usepackage{tikz}
\usepackage{xcolor}
\usetikzlibrary{calc}

\pgfplotsset{compat=newest}

\setcopyright{acmcopyright}
\copyrightyear{2020}
\acmYear{2020}
\acmDOI{}

\acmConference[ICCAD '20]{ICCAD '20: International Conference on Computer Aided Design}{November 02--05, 2020}{San Diego, CA}
\acmBooktitle{}
\acmPrice{}
\acmISBN{}

\makeatletter
\@printcopyrightfalse
\@printpermissionfalse
\@acmownedfalse
\@ACM@nonacmtrue
\setcopyright{none}
\makeatother

\begin{document}
\title[Lowering the $T$-depth of Quantum Circuits By Reducing the Multiplicative Depth Of Logic Networks]{Lowering the $T$-depth of Quantum Circuits By Reducing \\the Multiplicative Depth Of Logic Networks}
\author[T.~Haener and M.~Soeken]{Thomas Haener \qquad Mathias Soeken}
\affiliation{\institution{Microsoft Quantum, Switzerland}}

\begin{abstract}
  The multiplicative depth of a logic network over the gate basis $\{\land,
  \oplus, \neg\}$ is the largest number of $\land$ gates on any path from a
  primary input to a primary output in the network. We describe a dynamic
  programming based logic synthesis algorithm to reduce the multiplicative depth
  in logic networks.  It makes use of cut enumeration, tree balancing, and
  exclusive sum-of-products (ESOP) representations.  Our algorithm has
  applications to cryptography and quantum computing, as a reduction in the
  multiplicative depth directly translates to a lower $T$-depth of the
  corresponding quantum circuit.  Our experimental results show improvements in
  $T$-depth over state-of-the-art methods and over several hand-optimized quantum
  circuits for instances of AES, SHA, and floating-point arithmetic.
\end{abstract}

\maketitle

\section{Introduction}
Logic networks are the central data structure in logic optimization algorithms,
which have been widely applied for technology-independent optimization in
electronic design automation applications~\cite{TSAM19b,BHS90,Muroga93}.
Roughly speaking, the number of logic gates in a logic network corresponds to the
size of a physical implementation, while the number of logic levels corresponds
to its delay.

In recent years, the domain of applications for logic optimization has broadened
to also target areas such as cryptography~\cite{BMP13} and fault-tolerant
quantum computing (see, e.g., \cite{SPMH03,MS12,MS13,SRWM19}). Logic networks
are typically represented over a gate set consisting of 2-input AND gates,
2-input XOR gates, and inverters, called XOR-AND graphs (XAGs), in which only
the AND gates contribute to the cost functions.  The multiplicative complexity
(MC,~\cite{Schnorr88}) and the multiplicative depth (MD,~\cite{CAS17}) of a
Boolean function are two important theoretical metrics.  The multiplicative
complexity is the smallest number of AND gates necessary in any XAG that
represents the function. Similarly, the multiplicative depth of a function is
the smallest critical path (only considering AND gates) in any XAG that
represents the function. We also refer to the length of the critical path (only
considering AND gates) as AND-depth.

Multiplicative complexity and depth play important roles in cryptography
and fault-tolerant quantum computing.
A low multiplicative complexity corresponds to a higher
vulnerability to some cryptographic attacks.  In fault-tolerant quantum
computing, the multiplicative complexity provides an upper bound on the number
of expensive quantum operations as well as the number of qubits~\cite{MSC+19}.
Furthermore, the multiplicative depth corresponds to the execution time of a quantum
algorithm~\cite{MSRM20}. Computing the multiplicative complexity of a Boolean
function $f$ is expensive. It has been shown that no algorithm exists to compute
the multiplicative complexity that is polynomial in the size of the truth table
for $f$~\cite{Find14} if one-way functions~\cite{Levin03} exist.  We are not
aware of any theoretical results concerning the multiplicative depth.

We refer to the number of AND gates and the AND-depth of an XAG by \emph{MC/MD
of an XAG}, respectively. We use just MC or MD if it is clear from the context
whether we refer to the MC/MD of a Boolean function or to the MC/MD of a logic
network. We note that the latter provides an upper bound to the former.

Thus, many heuristics have been proposed that reduce the MC of an XAG (see,
e.g.,~\cite{BMP13,TSAM19,RJHK19,CCDE19,TSR+20}), aiming to arrive at tighter
upper bounds on the MC of the function being implemented. Similarly, some
heuristics have been proposed that aim to reduce the multiplicative
depth~\cite{CAS17,ACS20}.  In this paper, we introduce a logic synthesis
algorithm to reduce the MD of logic networks. Our algorithm is based on dynamic
programming and makes use of cut enumeration~\cite{CWD99}, tree
balancing~\cite{MBJK11}, as well as ESOP~\cite{Sasao93} and ESPP~\cite{IHKS04}
representations.

\textbf{Contributions.}\enspace We present a fully automatic logic synthesis
algorithm that reduces the multiplicative depth of logic networks.  We present
benchmarks demonstrating that our algorithm is capable of reducing the MD by up
to $3\times$ for depth-optimized logic networks and up to $9\times$
for MC-optimized logic networks. As a result, also the quantum
circuits derived from our depth-optimized networks feature depths that are
significantly smaller than state-of-the-art circuit designs. Crucially, these
improvements in depth are possible without increasing the number of qubits
significantly.

\section{Preliminaries}

\subsection{Logic networks}
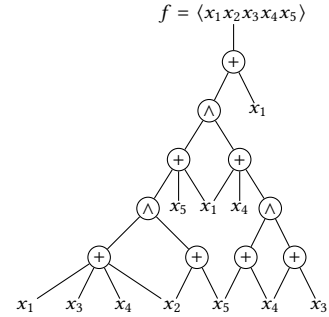
\begin{figure}
  \centering
  \begin{tikzpicture}[x=.65cm,y=0.65cm,font=\footnotesize]
    \begin{scope}[every node/.style={inner sep=.5pt}]
      \node at (0,0) (i1a) {$x_1$};
      \node at (1,0) (i3a) {$x_3$};
      \node at (2,0) (i4a) {$x_4$};
      \node at (3,0) (i2a) {$x_2$};
      \node at (4,0) (i5a) {$x_5$};
      \node at (5,0) (i4b) {$x_4$};
      \node at (6,0) (i3b) {$x_3$};
      \node at (3.125,2) (i5b) {$x_5$};
      \node at (3.75,2) (i1b) {$x_1$};
      \node at (4.375,2) (i4c) {$x_4$};
      \node at (4.75,4) (i1c) {$x_1$};
      \node at (4.25,6) (y) {$f = \langle x_1x_2x_3x_4x_5\rangle$};
    \end{scope}

    \begin{scope}[every node/.style={draw,circle,inner sep=0pt,minimum size=8.5pt}]
      \node at (1.5,1) (x1) {$+$};
      \node at (3.5,1) (x2) {$+$};
      \node at (4.5,1) (x3) {$+$};
      \node at (5.5,1) (x4) {$+$};
      \node at (2.5,2) (a1) {$\land$};
      \node at (5,2) (a2) {$\land$};
      \node at (3.125,3) (x5) {$+$};
      \node at (4.375,3) (x6) {$+$};
      \node at (3.75,4) (a3) {$\land$};
      \node at (4.25,5) (x7) {$+$};
    \end{scope}

    \draw (x1) -- (i1a) (x1) -- (i3a) (x1) -- (i4a) (x1) -- (i2a);
    \draw (x2) -- (i2a) (x2) -- (i5a);
    \draw (x3) -- (i5a) (x3) -- (i4b);
    \draw (x4) -- (i4b) (x4) -- (i3b);
    \draw (a1) -- (x1) (a1) -- (x2);
    \draw (a2) -- (x3) (a2) -- (x4);
    \draw (x5) -- (a1) (x5) -- (i5b) (x5) -- (i1b);
    \draw (x6) -- (i1b) (x6) -- (i4c) (x6) -- (a2);
    \draw (a3) -- (x5) (a3) -- (x6);
    \draw (x7) -- (a3) (x7) -- (i1c);
    \draw (x7) -- (y);
  \end{tikzpicture}
  \caption{XAG for the majority-of-5 function, with multiplicative complexity 3
    and multiplicative depth 2; for 2-input XOR gates are merged into
    multi-input XOR gates.}
  \label{fig:xag}
\end{figure}
In this work, we consider XOR-AND graphs (XAGs), which are logic networks
consisting of $2$-input AND gates, $2$-input XOR gates.  Such logic networks can
represent all $0$-preserving Boolean functions, i.e., functions $f$ for which
$f(0, \dots, 0) = 0$.  We are interested in logic networks that minimize the
maximum number of AND gates on any path from an input to an output as a primary
cost criteria, and the number of overall AND gates as a secondary cost criteria.
Functions $f$, which are not $0$-preserving, can be realized by finding an XAG
for $\bar f$ and then inverting the output.  Restricting to have inversions only
at the outputs does not affect the AND gates in the circuit, as all inner
inversions can be propagated to the outputs by only using XOR
gates~\cite{Schnorr88}.

Formally, we model an XAG for a single-output Boolean function $f$ over $n$
variables $x_1, \dots, x_n$ as a sequence of \emph{steps}, or gates,
\begin{equation}
  x_i = x_{j_{1i}} \circ_i x_{j_{2i}}
\end{equation}
for $n < i \le n+r$, and $\circ_i \in \{\oplus, \land\}$.  The values $1 \le
j_{1i} < j_{2i} < i$ point to primary inputs or previous steps in the network.
The function value is computed by the last step $f = x_{n+r}$.  This model is
readily extended to multi-output Boolean functions, by associating each output
function with some step in the network.  The \emph{logic level} of a primary
input or gate $i$ is defined as
\begin{equation}
  \label{eq:level}
  \ell_i =
  \begin{cases}
    0 & \text{if $i\le n$,} \\
    \max\{ \ell_{j_{1i}}, \ell_{j_{2i}} \} & \text{if $i > n$ and $\circ_i = \oplus$,} \\
    \max\{ \ell_{j_{1i}}, \ell_{j_{2i}} \} + 1 & \text{if $i > n$ and $\circ_i = \land$.}
  \end{cases}
\end{equation}
The depth of an XAG is $d = \max\{\ell_i\mid 1 \le i \le n+r\}$, the largest
level among all gates.  In other words, the logic level of a step is the
earliest possible time in which a step must be computed, if we aim at
parallelizing the evaluation of a logic network.  Similarly, we define the
\emph{reverse logic level} $\ell^r_i$ as the latest possible time in which step
$i$ must be computed while not increasing the depth of the logic network.

\begin{figure*}
  \footnotesize
  \centering
  \input{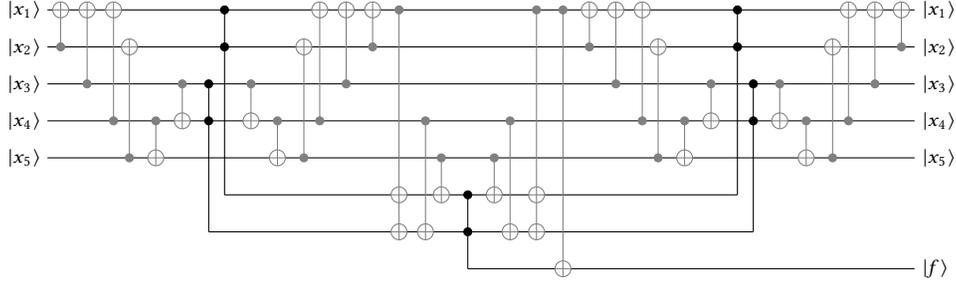}
  \caption{Quantum circuit for the majority-of-5 function, derived from the
  logic network in Fig.~\ref{fig:xag}.  Each of the computing AND gates, can
  be realized in $T$-depth 1 when using one helper qubit.  The circuit therefore
  has $T$-count 12, $T$-depth 2, and 11 qubits, incl.~qubits for function inputs
  and outputs.}
  \label{fig:qc}
\end{figure*}

\subsection{Cut enumeration}
Many logic optimization algorithms are based on applying local changes to small
subnetworks instead of considering the whole logic network at once. An important
family of single-rooted subnetworks are \emph{cuts}. Formally, a \emph{cut} $C$
of a step $i$ in a logic network is a set of steps, called \emph{leaves}, such
that (i) every path from step $i$ to a primary input visits at least one leaf,
and (ii) each leaf is contained in at least one path. Step $i$ is called the
\emph{root} of the cut and each cut represents a subgraph that includes the root
$i$ and some internal steps, and has the leaves as primary inputs. A cut is
$k$-feasible (referred to as $k$-cut), if $|C| \le k$, i.e., it has at most $k$
leaves.

Cut enumeration~\cite{CWD99} is an algorithm that computes all or a subset of
all $k$-cuts for each step in a network. It constructs a mapping
$\mathtt{CUTS}(i)$ that maps each step to a set of cuts using the following
recursive procedure:
\begin{equation}
  \label{eq:cuts}
  \mathtt{CUTS}(i) =
  \begin{cases}
    \{\{i\}\} & \text{if $i \le n$,} \\
    \{\{i\}\} \cup \bigcup \{C_1 \cup C_2 \mid & \\
    \qquad C_1 \in \mathtt{CUTS}(j_{1i}), & \\
    \qquad C_2 \in \mathtt{CUTS}(j_{2i}) & \\
    \qquad \text{s.t.}\,|C_1 \cup C_2| \le k\} & \text{otherwise.}
  \end{cases}
\end{equation}

Cuts $\{\{i\}\}$ for root $i$ are called \emph{trivial} cuts. Note that these
are essential, since otherwise the leaves of cuts can only be primary inputs. Cut
enumeration can also compute the function $\mathtt{FUNC}(i, C)$ represented by a
$C$ for root $i$, by assigning $\mathtt{FUNC}(i, \{i\}) = x_i$ for all trivial
cuts, and
\begin{equation}
  \mathtt{FUNC}(i, C) = \mathtt{FUNC}(j_{1i}, C_1) \circ_i \mathtt{FUNC}(j_{2i}, C_2)
\end{equation}
if $C$ was constructed using $C_1$ and $C_2$ in~\eqref{eq:cuts}.
Support-normalized truth tables are typically used to represent the cut
functions; e.g., truth tables for cut functions $x_1 \land x_3$ and $x_4 \land
x_9$, are both represented by the 4-bitstring $(1000)_2$.  To which variables
the truth table refers can be determined from the cut's leaves.

\subsection{Exclusive sum-of-products}
An ESOP for an $n$-variable Boolean function $f(x_1, \dots, x_n)$ has the form
\begin{equation}
f(x_1, \dots, x_n) = \bigoplus_{j=1}^m \left(x^{p_{1,j}}_1 \land \cdots \land
  x^{p_{n,j}}_n\right)
\end{equation}
for some $m$ and \textit{polarities} $p_{i,j}$, which take values from $0$ to
$2$. Their meaning is that $x_i^0 = \bar x_i$, $x_i^1 = x_i$, and $x_i^2 = 1$.
We call $x_i^0$ a negative literal, $x_i^1$ a positive literal, and $x_i^2$ an
empty literal. If $m = 0$, we define $f(x_1, \dots, x_n) = 0$.  The constant-1
function can be represented by an ESOP where $m=1$ and $p_{1,1} = \cdots =
p_{n,1} = 2$.

Each term $\left(x^{p_{1,j}}_1 \land \cdots \land x^{p_{n,j}}_n\right)$ is
called a \emph{cube} of \emph{degree} $d_j = |\{i \mid p_{i,j} \neq 2\}|$. It
can be regarded as an $(n - d_j)$-dimensional subcube of the $n$-dimensional
hypercube, in which the $2^n$ vertices correspond to all bitstrings of length
$n$.  We require that no cube occurs more than once in an ESOP.  The
\emph{degree} of the ESOP is $\max_{1\le j\le m} d_j$.

An ESOP in which $p_{i,j} \neq 0$ for all $1 \le i \le n$, $1 \le j \le m$ is
called the \emph{algebraic normal form} of $f$.  It is unique up to permutation
of the cubes.  The degree of the algebraic normal form is called the
\emph{algebraic degree} of $f$ and is a lower bound for the degree of any ESOP
for $f$.  An ESOP can be translated into the algebraic normal form by replacing
each cube with $2^l$ cubes in which all $l = |\{i \mid p_{i,j} = 0\}|$ negative
literals are replaced by all combinations of positive and empty literals.

Various exact and heuristic
algorithms~\cite{HP88,SB90,BS93,Sasao93,Drechsler99,MP01,SDP04,Papakonstantinou14,RESM20}
exist to find ESOPs for Boolean functions, where the primary cost function is
the number of cubes in the ESOP and the secondary cost function is the total
number of non-empty literals.  The positive of impact of ESOP expressions to our
work is mainly that they have a small depth, thereby having the potential to
reduce the multiplicative depth, however, they likely introduce a lot of AND
gates to express the cubes.  An ESOP optimization algorithm that targets the
number of literals as primary cost would therefore be a better fit for our
application.

\newcommand{\tikzCNOT}{%
\begin{tikzpicture}[baseline=(c)]
  \draw (0,0) -- (0,8pt);
  \fill (0,8pt) circle (1.1pt);
  \draw (0,0.08) circle (0.08);
  \draw (-0.08,0.08) -- ++(right:0.16);
  \coordinate (c) at (0,0.08);
\end{tikzpicture}
}
\newcommand{\tikzAND}{%
\begin{tikzpicture}[baseline=(c)]
  \draw (0,0) -- (0,8pt);
  \fill (0,8pt) circle (1.1pt);
  \fill (0,4pt) circle (1.1pt);
  \draw (0,0) -- (right:0.08);
  \coordinate (c) at (0,0.08);
\end{tikzpicture}
}
\newcommand{\tikzANDUC}{%
\begin{tikzpicture}[baseline=(c)]
  \draw (0,0) -- (0,8pt);
  \fill (0,8pt) circle (1.1pt);
  \fill (0,4pt) circle (1.1pt);
  \draw (0,0) -- (left:0.08);
  \coordinate (c) at (0,0.08);
\end{tikzpicture}
}

\subsection{Quantum computing}
A quantum computer contains quantum bits, so-called \textit{qubits}, to which
\textit{quantum gates} are applied in order to solve a computational task. It is
controlled by a classical computer running a quantum program, which consists of
both classical and quantum instructions: classical instructions are executed by
the (classical) host computer, and quantum instructions get sent to the quantum
co-processor for execution. In each computational step, the classical computer
decides on the sequence of quantum instructions to be executed on the
co-processor. Such sequences can be depicted as \textit{quantum circuits}. The
circuit diagram is read from left to right, with each horizontal line
representing a qubit, and quantum gates are represented as boxes/symbols on
these lines. Fig.~\ref{fig:qc} shows a quantum circuit that computes the
majority-of-5 function and is derived from the logic network in
Fig.~\ref{fig:xag}. The circuit consists of CNOT gates \tikzCNOT, AND gates
\tikzAND, as well as uncomputing AND gates \tikzANDUC.  CNOT gates act on two
qubits and compute the XOR of both qubit values onto the lower (target) qubit,
leaving the upper (control) qubit unchanged.  The AND gate computes a 1 on a
newly initialized target qubit, if and only if the two control qubits are 1. The
uncomputing AND gate expects that the target qubit is $1$ if and only if the
two control qubits are are 1, and releases the target qubit in a clean state
such that it can be used for subsequent computations.

In this paper, we target quantum computing running a protocol for
fault-tolerance, which is necessary to run quantum algorithms with more than a
few thousand operations, e.g., for chemistry simulations of practical
interest~\cite{RWS+17}. In this setting, the focus of circuit optimization
shifts away from two-qubit gates (e.g., for NISQ devices~\cite{Preskill18})
toward gates that require distillation. In particular, when the surface code is
used, the so-called $T$-gate incurs a large overhead~\cite{AMMR13,OC17}. In
fault-tolerant quantum computing, the cost of CNOTs are typically neglected. The
AND gate has a $T$-count of $4$ and a $T$-depth of 1, if one additional helper
qubit is used for its implementation~\cite{Jones13} (otherwise, it can be
implemented with a $T$-depth of 2 without the use of a helper qubit).  The
uncomputing AND gate requires no $T$-gates.

Previous work~\cite{MSC+19} focused on reducing the number of costly $T$-gates.
Instead, we aim to shorten the time to solution by reducing the $T$-depth
instead.

\section{Multiplicative depth reduction}

In this section, we introduce various methods that reduce the multiplicative depth of logic networks. Then, we present a procedure to map these networks to quantum circuits while maintaining depth improvements.

\subsection{Cut-based balancing}
\begin{algorithm}
  \small
  \begin{algorithmic}
    \FOR{$i = 1, \ldots, n$}
      \STATE $\mathtt{BEST}(i) \gets i$
    \ENDFOR
    \FOR{$i = n + 1, \dots, n + r$}
      \STATE $j_{\mathrm{best}} \gets \Lambda$
      \STATE $\ell_{j_{\mathrm{best}}} \gets \infty$
      \FOR{$C \in \mathtt{CUTS}(i)$ s.t. $|C|>1$}
        \STATE $\{l_1, \dots, l_k\} \gets C$
        \STATE $j \gets \mathit{balance}(\mathtt{FUNC}(i, C), \mathtt{BEST}(l_1), \dots, \mathtt{BEST}(l_k))$
        \IF{$\ell_j < \ell_{j_{\mathrm{best}}}$}
          \STATE $j_{\mathrm{best}} \gets j$
        \ENDIF
      \ENDFOR
      \STATE $\mathtt{BEST}(i) \gets j_{\mathrm{best}}$
    \ENDFOR
    \RETURN $\mathtt{BEST}(n+r)$
  \end{algorithmic}
  \caption{Generic cut-based balancing}
  \label{alg:balancing}
\end{algorithm}
Algorithm~\ref{alg:balancing} describes a generic balancing algorithm based on
dynamic programming and cut enumeration inspired by~\cite{MBJK11}.  It takes as
input a logic network for an $n$-variable Boolean function with $r$ steps and
returns a new depth-optimized logic network.  Traversing all steps $i$ in
topological order, it computes depth-optimized candidates for each cut $C$ of
$i$, and stores the best candidate in a mapping $\mathtt{BEST}(i)$.  The output
of the depth-optimized network is $\mathtt{BEST}(n+r)$ after all steps have been
visited.  For each cut $C$ of step $i$, the algorithm tries to resynthesize the
cut function $\mathtt{FUNC}(i, C)$ with the target to reduce the level of step
$i$.  For this purpose, it assumes the best candidates for the cut's leaves.

The algorithm uses a $\mathit{balance}$ function to resynthesize the cut
function.  It is therefore generic and can be customized by applying various
resynthesis procedures.  One possible resynthesis procedure is presented
in~\cite{MBJK11}.  It computes a sum-of-products (SOP) representation for the
cut function and then translates each term in the SOP into a weight-balanced
tree of AND gates, as well as all terms into a weight-balanced tree of OR gates.
Our work adapts this method by using an ESOP representation instead, where the
outer XOR operations do not contribute to the logic network's multiplicative
depth.

\subsection{ESOP balancing}
In this section we discuss a rebalancing algorithm based on ESOP forms, which
can be used in Algorithm~\ref{alg:balancing}.  ESOP forms offer a potentially
low-depth implementation as an XAG.  For the sake of a simpler description of
the algorithm, we assume that the ESOP form is given in algebraic normal form,
however, in the implementation we consider ESOP forms that also contain negative
literals, since they allow for a more compact representation.

Given a $k$-cut $C = \{l_1, \dots, l_k\}$ of root $i$ with cut function
\[
	\mathtt{FUNC}(i, C) = f(\hat x_1, \dots, \hat x_k),
\]
where $\hat x_i =
x_{\mathtt{BEST}(l_i)}$ with corresponding level $\hat\ell_i$.  If we are given
an ESOP for $f$ with $m$ cubes, then each cube is translated into a tree of
2-input AND gates that is balanced with respect to the leaf levels.  Then all
outputs of these AND-trees are combined by a tree of 2-input XOR gates, which
does not add to the multiplicative depth.

\begin{algorithm}
  \small
  \begin{algorithmic}
    \REQUIRE Product term $p_{1,j}, \dots, p_{k,j}$, variables
      $\hat x_1, \dots, \hat x_k$ at levels $\hat\ell_1, \dots, \hat\ell_k$
    \STATE let $Q$ be a priority queue of steps, ordered by the steps' level in ascending order
    \FOR{$i=1, \dots, k$ s.t. $p_{i,j}=1$}
      \STATE $\mathit{push}(Q, \hat x_i)$
    \ENDFOR
    \WHILE{$|Q| > 1$}
      \STATE $u \gets \mathit{pop}(Q)$
      \STATE $v \gets \mathit{pop}(Q)$
      \STATE $\mathit{push}(Q, u \land v)$
    \ENDWHILE
    \RETURN $\mathit{pop}(Q)$
  \end{algorithmic}
  \caption{Tree balancing computation of product term~$j$}
  \label{alg:tree-balancing}
\end{algorithm}

The algorithm to balance a non-constant ESOP cube with respect to the leaf
levels is described in Algorithm~\ref{alg:tree-balancing}.  First all non-empty
literals are inserted into a priority queue $Q$ according to their levels in
ascending order.  Then as long as the queue has more than one element, the two
top-most elements are popped from the queue and merged with an AND gate.  The
resulting step is then pushed back into the queue, taking the level of the step
into account for the ordering.

\begin{table*}
  \bgroup
  \caption{Experimental results for applying ESOP-balancing}
  \label{tbl:esop}
  \catcode`\*=\active\def*{\bf}%
  \small%
  \begin{tabularx}{\linewidth}{X rrr r@{}>{\;(}r<{)}r@{}>{\;(}r<{)}r r@{}>{\;(}r<{)}r@{}>{\;(}r<{)}r}
    \toprule
    Benchmark &
    \multicolumn{3}{c}{State-of-the-art~\cite{CAS17,ACS20}} &
    \multicolumn{5}{c}{Min.~MC baseline} &
    \multicolumn{5}{c}{Min.~depth baseline} \\
    \midrule
    &
    {\scriptsize MC} & {\scriptsize MD} & {\scriptsize Run-time} &
    \multicolumn{2}{c}{\scriptsize MC (before)} & \multicolumn{2}{c}{\scriptsize MD (before)} & {\scriptsize Run-time} &
    \multicolumn{2}{c}{\scriptsize MC (before)} & \multicolumn{2}{c}{\scriptsize MD (before)} & {\scriptsize Run-time} \\[4pt]
    \multicolumn{14}{l}{\textit{Arithmetic functions~\cite{AGM15}}} \\[3pt]
    adder      &   16378 &    *9 &    125.00 &      481 &   128 &   34 &  128 &    0.15 &   2761 &   1742 &   12 &   14 &   10.13 \\
    bar        &    4193 &    10 &      0.70 &     1303 &   832 &   *4 &    7 &    0.33 &   3516 &   3334 &    8 &   11 &    2.42 \\
    div        &  190855 &   532 &   3731.00 &   158795 &  5288 &  973 & 2243 &   26.18 & 120327 & 120327 & *523 &  620 &  541.33 \\
    hyp        &  135433 & 15230 & 172000.00 &   120765 & 56635 & 4428 & 8784 &  166.07 & 780220 & 417567 &*1287 & 1558 &  324.31 \\
    log2       &   31573 &   129 &     94.00 &    34133 & 10906 & *104 &  201 &  778.78 &  83177 &  33951 &  114 &  171 &  130.33 \\
    max        &    7666 &    26 &     14.50 &     3839 &   890 &   93 &  252 &    1.81 &   8368 &   4027 &  *25 &   28 &    4.17 \\
    multiplier &   23059 &    57 &     30.73 &    15138 &  7653 &   65 &  149 &   13.50 &  39628 &  28331 &  *56 &   86 &   77.72 \\
    sin        &    5507 &    74 &      4.50 &     6822 &  2603 &   62 &  105 &    9.45 &  14067 &   6424 &  *61 &   89 &   58.87 \\
    sqrt       &  321555 &  2084 & 107814.00 &    71587 &  5381 &  951 & 2167 &   45.75 & 185061 &  65762 & *769 &  936 &  290.40 \\
    square     &   11306 &    26 &     12.50 &     6348 &  4672 &   59 &  155 &    8.41 &  10777 &  14570 &  *20 &   36 &   38.75 \\[4pt]
    \multicolumn{14}{l}{\textit{Random control~\cite{AGM15}}} \\[3pt]
    arbiter    &    5183 &   *10 &     43.00 &     3128 &  1174 &   13 &   50 &    2.10 &   7276 &   6205 &   11 &   12 &    1.35 \\
    cavlc      &     667 &     9 &      0.00 &      447 &   394 &   *7 &   11 &    1.15 &    564 &    576 &    8 &   10 &    0.45 \\
    ctrl       &     109 &     5 &      0.00 &       54 &    45 &   *4 &    5 &    0.10 &     77 &     80 &    4 &    8 &    0.06 \\
    dec        &     304 &     3 &      0.00 &      328 &   328 &    3 &    3 &    0.08 &    292 &    292 &   *3 &    3 &    0.02 \\
    i2c        &    1213 &     7 &      0.10 &      816 &   557 &   *7 &   11 &    0.87 &   1122 &   1007 &    7 &    8 &    0.37 \\
    int2float  &     216 &     7 &      0.00 &      104 &    85 &   *6 &   11 &    0.87 &    184 &    190 &    7 &    8 &    0.13 \\
    mem\_ctrl  &   54816 &    40 &     85.00 &     9983 &  4695 &  *14 &   39 &   17.56 &  78044 &  37519 &   35 &   41 &   20.37 \\
    priority   &     876 &   102 &      0.50 &      442 &   323 &   11 &   95 &    1.08 &    522 &    479 &  *10 &   13 &    0.28 \\
    router     &     198 &    11 &      0.00 &      116 &    93 &   *8 &   13 &    0.10 &    227 &    196 &   10 &   12 &    0.19 \\
    voter      &    4288 &    30 &    112.42 &     7335 &  4257 &   26 &   40 &   31.95 &   3255 &   6716 &  *17 &   48 &    6.14 \\[4pt]
    \multicolumn{14}{l}{\textit{Cryptographic functions~\cite{AAM+}}} \\[3pt]
    AES-128    &         &       &           &    8400 &   6400 &  *50 &   60 &    5.49 &  33953 &  85547 &   80 &  299 &   65.52 \\
    AES-192    &         &       &           &    9408 &   7168 &  *60 &   72 &    5.98 &  39533 &  96979 &   99 &  359 &   55.39 \\
    AES-256    &         &       &           &   11592 &   8832 &  *70 &   84 &    7.49 &  53775 & 120627 &  123 &  417 &   90.26 \\
    Keccak-f   &         &       &           &   38400 &  38400 &  *24 &   24 & ---     &  38630 & 567395 &   28 &  266 &  129.00 \\
    SHA-256    &         &       &           &   22573 &  22573 & 1607 & 1607 & ---     & 450447 & 296951 &*1519 & 1936 &  247.16 \\
    SHA-512    &         &       &           &   57947 &  57947 & 3303 & 3303 & ---     &1988586 & 831166 &*2383 & 2894 & 1489.64 \\[4pt]
    \multicolumn{14}{l}{\textit{IEEE floating-point operations~\cite{AAM+}}} \\[3pt]
    FP-add     &         &       &           &   16721 &   5384 &   96 &  235 &    9.93 &  27541 &  15879 &  *64 &   83 &   15.42 \\
    FP-div     &         &       &           & 3829444 &  82265 & 1646 & 3619 & 2994.35 & 732932 & 200112 & *885 & 1157 &  400.12 \\
    FP-eq      &         &       &           &     315 &    315 &    9 &    9 & ---     &    220 &    336 &   *9 &   10 &    0.02 \\
    FP-f2i     &         &       &           &    3290 &   1467 &   24 &   94 &    3.01 &   3405 &   2881 &  *21 &   29 &    3.41 \\
    FP-mul     &         &       &           &   23886 &  19614 &   92 &  129 &   14.78 &  62254 &  47213 &  *87 &  140 &   54.51 \\
    FP-sqrt    &         &       &           & 4946577 &  91499 & 3763 & 6507 & 2981.18 & 893849 & 264130 &*1877 & 2374 &  506.28 \\
    \bottomrule
  \end{tabularx}
  \egroup
\end{table*}

\subsection{ESPP optimization}
An exclusive sum-of-pseudoproducts (ESPP~\cite{IHKS04}) for an $n$-variable
Boolean function $f(x_1, \dots, x_n)$ has the form
\begin{equation}
  \label{eq:espp}
  f(x_1, \dots, x_n) = \bigoplus_{j=1}^m
  \left(
    L_0^{p_{0,j}} \land \cdots \land L_{2^n-1}^{p_{2^{n-1},j}}
  \right)
\end{equation}
where $L_i = b_1x_1 \oplus \cdots \oplus b_nx_n$ when $i = (b_n \ldots b_1)_2$
is the linear function (or parity function) that contains variables according to
the positions of 1s in the binary expansion of $i$.  The polarity variables
$p_{i,j}$ play the same role as defined for ESOP forms, i.e., the parity
function $L_i$ in term $j$ is negated if $p_{i,j} = 0$, used as is if $p_{i,j} =
1$, and omitted if $p_{i,j} = 2$.  The terms in~\eqref{eq:espp} are called
\emph{pseudoproducts}~\cite{LP99}.  Note that each ESOP is also an ESPP, but an
ESPP is only an ESOP if $(p_{i,j} \neq 2) \rightarrow (\nu(i) = 1)$ (where $\nu(i)$ is the sideways sum of $i$, i.e., the number of 1s in its binary expansion).

The authors presented an exhaustive search algorithm to find small ESPPs
in~\cite{IHKS04}, and some theoretical investigations on the form have been
conducted~\cite{Selezneva14}.  However, to the best of our knowledge no
efficient heuristic optimization algorithm for ESPPs has been presented.

We implemented a simple heuristic Greedy minimization algorithm to minimize the
number of terms in an ESPP.  The algorithm iteratively merges cubes to increase
the use of linear functions as cube literals, thereby minimizing the number of
AND operations.  The algorithm starts with an initial ESPP form that corresponds
to an ESOP form, extracted from a cut function.  It then checks whether there
exists two distinct terms with indices $j_1$ and $j_2$ such that there exist two
indices $0\le i_1, i_2 < 2^n$ such that $p_{i_1,j_1} = p_{i_2,j_2} = 2$ but
$p_{i_1,j_2} \neq 2$ and $p_{i_2,j_1} \neq 2$, and for all other indices $i
\notin \{i_1,i_2\}$, it holds that $p_{i,j_1} = p_{i,j_2}$. Then, the two terms
can be combined into a single term $j$, with $p_{i,j} = p_{i,j_1}$ for all $i
\notin \{i_1,i_2\}$ and
\begin{equation}
  p_{i_1 \oplus i_2, j} = [p_{i_1,j_2} = p_{i_2,j_1}]
\end{equation}
if $p_{i_1 \oplus i_2, j_1} \in \{2, [p_{i_1,j_2} = p_{i_2,j_1}]\}$.  If $p_{i_1
\oplus i_2, j_1} = 1 - [p_{i_1,j_2} = p_{i_2,j_1}]$, the two terms cancel and
can be removed without adding another term to the ESPP.  We iterate this
procedure until no more such two terms can be found.  In our implementation,
empty parity functions are not explicitly stored, and therefore this procedure
can be efficiently implemented.

\begin{table}[t]
  \caption{Estimates for $T$-depth optimized quantum circuits obtained from depth-optimized XAGs.  We report quantum circuits that achieve the smallest number of qubits (first row) and the lowest $T$-depth (second row) over all $T$-depth optimized circuits.}
  \label{tbl:quantum-count}
  \def\tabcolsep{4pt}
  \small
  \begin{tabularx}{\linewidth}{lrrrX}
    \toprule
    Benchmark & T-count & T-depth & Qubits & Instance \\
    \midrule
    \multicolumn{5}{l}{\textit{Cryptographic functions}} \\[3pt]
    AES-128         & 25600   & 60     & 7324    & Min.~MC baseline (ASAP)    \\
    AES-128         & 33600   & 50     & 9384    & Min.~MC opt (ASAP)         \\
    AES-192         & 28672   & 72     & 8156    & Min.~MC baseline (ASAP)    \\
    AES-192         & 37632   & 60     & 10456   & Min.~MC opt (ASAP)         \\
    AES-256         & 35328   & 84     & 9884    & Min.~MC baseline (ASAP)    \\
    AES-256         & 46368   & 70     & 12704   & Min.~MC opt (ASAP)         \\
    Keccak-f        & 153600  & 24     & 46400   & Min.~MC baseline (ASAP)    \\
    SHA-256         & 90292   & 1607   & 23684   & Min.~MC baseline (ASAP)    \\
    SHA-256         & 1801788 & 1519   & 458974  & Min.~depth opt (ASAP)      \\
    SHA-512         & 231788  & 3303   & 60448   & Min.~MC baseline (ASAP)    \\
    SHA-512         & 7954344 & 2383   & 2008595 & Min.~depth opt (ASAP)      \\[4pt]
    \multicolumn{5}{l}{\textit{IEEE floating-point operations}} \\[3pt]
    FP-add          & 21384   & 235    & 5969    & Min.~MC baseline (ALAP)    \\
    FP-add          & 100832  & 64     & 28154   & Min.~depth opt (ASAP)      \\
    FP-div          & 290848  & 3604   & 81066   & Min.~MC baseline (ALAP)    \\
    FP-div          & 3054524 & 885    & 792188  & Min.~depth opt (ASAP)      \\
    FP-eq           & 880     & 9      & 655     & Min.~depth opt (ALAP)      \\
    FP-eq           & 1260    & 9      & 976     & Min.~MC baseline (ASAP)    \\
    FP-f2i          & 5832    & 94     & 1821    & Min.~MC baseline (ALAP)    \\
    FP-f2i          & 13620   & 21     & 4846    & Min.~depth opt (ASAP)      \\
    FP-mul          & 76368   & 118    & 26890   & Min.~MC baseline (ALAP)    \\
    FP-mul          & 249052  & 87     & 69347   & Min.~depth opt (ASAP)      \\
    FP-sqrt         & 315924  & 6498   & 84017   & Min.~MC baseline (ALAP)    \\
    FP-sqrt         & 3575396 & 1877   & 901087  & Min.~depth opt (ASAP)      \\
    \bottomrule
  \end{tabularx}
\end{table}

\subsection{Mapping to quantum circuit}
Given a logic network over the gate basis $\{\land,\oplus,\neg\}$, it is
straightforward to generate a quantum circuit that computes the same function:
Each $\land$ node in the network can be mapped to a Toffoli that writes the
output into an extra qubit starting in $\ket0$; each $\oplus$ and $\neg$ node
can be computed inplace using a (controlled) NOT gate~\cite{MSC+19}.

While the resulting quantum circuit computes the same function, a significant
amount of parallelism is lost due to input-dependencies. As a remedy, we copy
the inputs of those gates that can be executed in parallel, thus removing these
dependencies~\cite{MSRM20}.

\section{Experimental Results}
\begin{figure*}[t]
  \begin{tikzpicture}[font=\footnotesize]
    \pgfplotsset{%
      every axis/.append style={%
        scale only axis,
        width=.26\linewidth,
        anchor=north west,
        xlabel=Qubits,
        ylabel=$T$-depth,
        ylabel shift=-4pt,
        xlabel shift=-4pt,
        ticklabel style={font=\tiny},
        axis background/.append style={fill=gray!5!white},
        cycle list={%
          {teal!80!black,only marks,mark=*,every mark/.append style={fill=teal}},
          {orange!80!black,only marks,mark=*,every mark/.append style={fill=orange}},
          {red!70!white!80!black,only marks,mark=*,every mark/.append style={fill=red!70!white}},
          {lime!80!black!80!black,only marks,mark=*,every mark/.append style={fill=lime!80!black}},
          {yellow!60!black!80!black,only marks,mark=*,every mark/.append style={fill=yellow!60!black}},
          {red!70!black!80!black,only marks,mark=*,every mark/.append style={fill=red!70!black}},
          {blue,only marks,mark=pentagon*,every mark/.append style={fill=blue!80!black}}
        },
      }
    }
    \tikzset{%
      jnrv19 arrow/.style={->,lime!80!black,dash pattern=on 2pt off 1pt}
    }
    \begin{loglogaxis}[title=AES-128,name=aes128]
      \addplot coordinates {(984,33264) (3017,14784) (2208,3780) (7148,1680) (6654,2160) (21854,960)};
      \label{plt:khj18}
      \addplot coordinates {(984,50688)};
      \label{plt:glrs16}
      \addplot coordinates {(864,7520)};
      \label{plt:lps20}
      \addplot coordinates {(1785,120) (2937,120)};
      \label{plt:jnrv19}
      \pgfplotsset{cycle list shift=2}
      \addplot coordinates {(9384, 50) (7324, 60)};
      \label{plt:ours}
    \end{loglogaxis}
    \begin{loglogaxis}[title=AES-192,name=aes192,xshift=.08\linewidth,at={(aes128.north east)}]
      \pgfplotsset{cycle list shift=1}
      \addplot coordinates {(1112,44352)};
      \addplot coordinates {(896,6560)};
      \addplot coordinates {(2105,120) (3513,120)};
      \pgfplotsset{cycle list shift=3}
      \addplot coordinates {(10456, 60) (8156, 72)};
    \end{loglogaxis}
    \begin{loglogaxis}[title=AES-256,name=aes256,xshift=.08\linewidth,at={(aes192.north east)}]
      \pgfplotsset{cycle list shift=1}
      \addplot coordinates {(1336,1505280)};
      \addplot coordinates {(1232,8640)};
      \addplot coordinates {(2425,126) (4089,126)};
      \pgfplotsset{cycle list shift=3}
      \addplot coordinates {(12704, 70) (9884, 84)};
    \end{loglogaxis}
    \begin{loglogaxis}[title=SHA-256,name=sha256,yshift=-18pt,at={(aes128.below south west)}]
      \addplot coordinates {(801,109104) (853,39840) (834,82752) (938,30336)};
      \pgfplotsset{cycle list shift=3}
      \addplot coordinates {(2402,70400)};
      \label{plt:amg+16}
      \pgfplotsset{cycle list shift=4}
      \addplot coordinates {(458974, 1519) (349079, 1522) (248256, 1543) (23684, 1607)};
    \end{loglogaxis}
    \begin{loglogaxis}[title=Keccak-f,name=keccak,xshift=.08\linewidth,at={(sha256.north east)}]
      \pgfplotsset{cycle list shift=4}
      \addplot coordinates {(24640,33)};
      \pgfplotsset{cycle list shift=5}
      \addplot coordinates {(46400, 24)};
    \end{loglogaxis}
    \begin{loglogaxis}[title=FP-add,name=fpadd,yshift=-18pt,at={(sha256.below south west)}]
      \pgfplotsset{cycle list shift=5}
      \addplot coordinates {(268,26348)};
      \label{plt:hsrs18}
      \addplot coordinates {(26734, 64) (20538, 65) (17065, 83) (14018, 98) (10562, 122) (5969, 235)};
    \end{loglogaxis}
    \begin{loglogaxis}[title=FP-mul,name=fpmul,xshift=.08\linewidth,at={(fpadd.north east)}]
      \pgfplotsset{cycle list shift=5}
      \addplot coordinates {(315,52116)};
      \label{plt:hsrs18}
      \addplot coordinates {(67146, 87) (56404, 88) (43565, 91) (32450, 92) (30504, 99) (26890, 118)};
    \end{loglogaxis}
    \coordinate (legend) at ($(current bounding box.north east)!2/3!(current bounding box.south east)$);
    \node[left,align=left] at (legend) {%
      \ref{plt:khj18}~Kim, Han, Jeong~\cite{KHJ18}\\[2pt]
      \ref{plt:glrs16}~Grassl, Langenberg, Roetteler,\\
      \phantom{\ref{plt:glrs16}}~Steinwandt~\cite{GLRS16}\\[2pt]
      \ref{plt:lps20}~Langenberg, Pham, Steindwandt~\cite{LPS20}\\[2pt]
      \ref{plt:jnrv19}~Jaques, Naehrig, Roetteler, Virdia~\cite{JNRV19}\\[2pt]
      \ref{plt:amg+16}~Amy, Matteo, Gheorghiu, Mosca, \\
      \phantom{\ref{plt:amg+16}}~Parent, Schanck~\cite{AMG+16}\\[2pt]
      \ref{plt:hsrs18}~Haener, Soeken, Roetteler, Svore~\cite{HSRS18}\\[2pt]
      \ref{plt:ours}~This work
    };
  \end{tikzpicture}
  \caption{These plots contain various resource estimates that can be found in the literature, together with all Pareto-optimal results for our approach that we obtained during the experimental evaluation.  These include also results from intermediate optimization steps.}
  \label{fig:estimates}
\end{figure*}
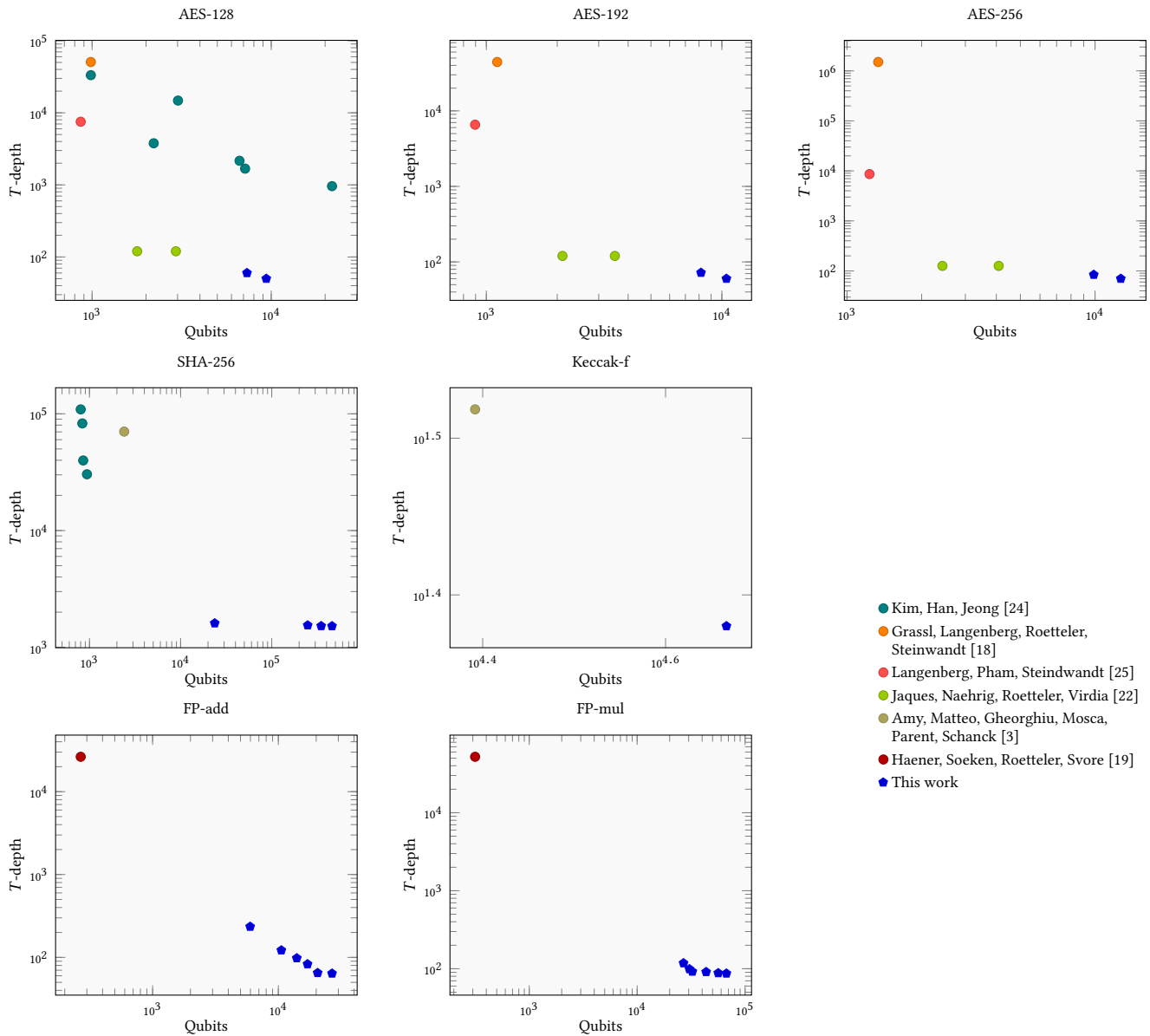

\begin{table}
  \caption{Estimates from related work}
  \label{tbl:other}
  \small
  \begin{tabularx}{\linewidth}{Xrrrr}
    \toprule
    Benchmark & $T$-count & $T$-depth & Qubits & Comment \\
    \midrule
    AES-128~\cite{KHJ18}   &     --- &   960 & 21854            & $^1$ \\[4pt]
    AES-128~\cite{GLRS16}  & 1060864 & 50688 &   984\rlap{$^*$} & \\
    AES-192~\cite{GLRS16}  & 1204224 & 44352 &  1112\rlap{$^*$} & \\
    AES-256~\cite{GLRS16}  & 1505280 & 59904 &  1336\rlap{$^*$} & \\[4pt]
    AES-128~\cite{LPS20}   &  118580 &  7520 &   864\rlap{$^*$} & $^2$ \\
    AES-192~\cite{LPS20}   &  137060 &  6560 &   896\rlap{$^*$} & $^2$\\
    AES-256~\cite{LPS20}   &  166320 &  8640 &  1232\rlap{$^*$} & $^2$\\[4pt]
    AES-128~\cite{JNRV19}  &   54400 &   120 &  1785\rlap{$^*$} & \\
    AES-192~\cite{JNRV19}  &   60928 &   120 &  2105\rlap{$^*$} & \\
    AES-256~\cite{JNRV19}  &   75072 &   126 &  2425\rlap{$^*$} & \\[4pt]
    Keccak-f~\cite{AMG+16} &   24640 &    33 &  3200\rlap{$^*$} & \\[4pt]
    SHA-256~\cite{KHJ18}   &     --- & 30336 &   938\rlap{$^*$} & $^1$ \\[4pt]
    SHA-256~\cite{AMG+16}  &  228992 & 70400 &  2402\rlap{$^*$} & \\[4pt]
    FP-add~\cite{HSRS18}   &   26348 &  7224 &   268\rlap{$^*$} & $^3$ \\
    FP-mul~\cite{HSRS18}   &  122752 & 52116 &   315\rlap{$^*$} & $^3$ \\
    \bottomrule
  \end{tabularx}

  \begin{flushleft}
  \medskip A $^*$ indicates that this value is better compared to the best value
  reported in Table~\ref{tbl:quantum-count}.

  \smallskip $^1$ Authors report no Toffoli-count or $T$-count; $T$-depth is
  derived from reported Toffoli-depth by multiplication with 3~\cite{AMMR13};
  authors report six different candidates, from which we picked the one with the
  best $T$-depth.

  \smallskip $^2$ $T$-count and $T$-depth are derived from reported
  Toffoli-count and Toffoli-depth in the paper.

  \smallskip $^3$ The floating-point designs in the paper are not IEEE-compliant
  and do not account for special cases or denormalized numbers.
  \end{flushleft}
\end{table}

We use various arithmetic and random-control functions from~\cite{AGM15} as well
as cryptographic functions and IEEE floating-point operations~\cite{AAM+} as
benchmarks for our algorithm.  Our algorithm has been implemented in C++ on top
of the EPFL logic synthesis libraries~\cite{SRHM18}.  All experiments were run
on a Microsoft Azure virtual machine, on a general purpose Standard D8s v3 size
configuration, running on an Intel Xeon Platinum 8171M 2.40GHz CPU with 32 GiB
memory and Ubuntu 18.04.

We choose two different baselines as starting points, heavily optimized XAGs for
low MC (Min.~MC baseline) and heavily optimized AIGs (And-inverter graphs) for
low (general) logic network depth (Min.~depth baseline).  The Min.~MC baseline
is obtained using the MC optimization algorithm in~\cite{TSR+20}.\footnote{The
cryptographic and floating-point operations were not further optimized, as they
are already optimized for MC.} The Min.~MC baseline is obtained by calling the
ABC~\cite{BM10} optimization scripts \texttt{resyn2rs} (depth-preserving size
optimization~\cite{MCB06,MB06}), followed by \texttt{if -K 6 -y} (AIG depth
optimization~\cite{YWM12}), followed by another round of \texttt{resyn2rs}, each
run until depth is no longer improved.

\subsection{Multiplicative-depth optimization}
As a first step, we apply ESOP-balancing with a cut size of 6 and
exorcism~\cite{MP01} to obtain ESOPs for the cut functions to the chosen
benchmarks for both baselines.  We call the algorithm repeatedly until no
further reduction in the multiplicative depth can be obtained.  We report the
results in Table~\ref{tbl:esop}. For the EPFL benchmarks we list the currently
best-known results for multiplicative depth obtained from the state-of-the-art
multiplicative depth optimization approach in~\cite{CAS17,ACS20}. That approach
has not been applied to the cryptographic and floating-point operations.  For
each baseline we list MC and MD after optimization together with the initial
values in parentheses, as well as runtime in seconds.  The result with the
lowest MD is highlighted in bold; in case of a tie we compare MC as a second cost
metric.

Our algorithm can improve the best-known results in 18 out of 20 cases.  For the
arithmetic functions, the largest MD reductions were obtained when applying our
approach to the Min.~depth baseline, whereas for the random control functions,
the Min.~MC baseline turns out to be the better starting point.  Note that in
some cases (e.g., \emph{hyp} and \emph{priority}) we obtain a $10\times$
improvement over the state of the art.  For the cryptographic and floating-point
functions, we can improve the MD compared to both baselines for all benchmarks
except for \emph{Keccak-f}.  Because we use heavily-optimized networks as the
baseline, we do not expect large gains for cryptographic functions, especially
since MC and MD are important quantities in cryptography. In contrast, we find
depth-reductions of up to $3\times$ for floating-point operations (e.g.,
\emph{FP-add} and \emph{FP-f2i}) with only moderate increases in MC.

\subsection{$T$-depth optimization}
In a second step, we map our depth-optimized XAGs to quantum circuits using two
straightforward heuristics for upper-bounding the number of qubits: the \emph{as
soon as possible} (ASAP) heuristic computes all AND gates in parallel that have
the same logic level and the \emph{as late as possible} (ALAP) heuristic
computes all AND gates in parallel that have the same reverse logic level. We
present the resulting $T$-counts, $T$-depths, and qubit estimates in
Table~\ref{tbl:quantum-count}. For each cryptographic function and
floating-point operation, we report the two quantum circuits with the fewest
number of qubits (first row) and the lowest $T$-depth (second row).  The
corresponding XAG and heuristic (ASAP or ALAP) is given in the last column.
Compared to state-of-the-art, manually-crafted quantum circuit designs, we
achieve significant reductions in depth without dramatically increasing the
qubit requirements. A comparison of our automatically-generated designs to a
variety of state-of-the-art circuits for several cryptographic and
floating-point functions is given in Fig.~\ref{fig:estimates} with best
$T$-depth state-of-the-art implementations explicitly reported in
Table~\ref{tbl:other}.

We note that, in addition to reduced circuit depths compared to the state of the
art, our approach has the clear advantage that it is completely automatic. This
stands in stark contrast to the circuits found in the literature, since those
are manual designs that were not created by the push of a button.

\section{Conclusions}
In this work we presented dynamic programming algorithm to minimize the
multiplicative depth of XAGs that makes use of cut enumeration, tree balancing,
as well as ESOP and ESPP representations.  We can report significant improvement
to the state-of-the-art MD optimization algorithms in~\cite{CAS17,ACS20}. We
used our algorithm to find fault-tolerant quantum implementations of various
cryptographic and floating-point operations that improve the $T$-depth over
state-of-the-art manual designed quantum circuits.

The adoption of SOP-based balancing for Boolean logic networks to ESOP-balancing
for XAGs in order to reduce the multiplicative depth worked very well, since the
XOR gates corresponding to the outer XOR operator of the ESOP forms does not
contribute to the depth.  We plan to investigate how this change in the cost
function and underlying logic representation may benefit from alternative depth
optimization algorithms such MUX-based optimization~\cite{BHLT90,MBJ10}),
generalized select transform algorithms~\cite{MBSS91,SHM+94}, or BDD-based
techniques~\cite{CCW07,CM10}.

We presented a post-optimization algorithm for ESOPs based on ESPPs, a
generalization of ESOPs.  XP$^2$ forms are a generalization of ESPPs and a
minimization algorithm for such forms has been presented in~\cite{VBI08}.  We
expect that such forms can help to further reduce the number of AND gates in the
rebalancing step of our algorithm without increasing the multiplicative depth.

In classical logic synthesis optimization flows, it is customary to interleave
depth-optimization algorithms with size-optimization algorithms to obtain good
trade-off points.  We plan to adopt heuristic MC optimization algorithms to be
depth-preserving, i.e., allowing the minimization of AND gates only if the
multiplicative depth does not increase.  This allows to reduce the $T$-count and
qubit count in corresponding quantum circuits without increasing the $T$-depth.

\bibliographystyle{ACM-Reference-Format}
\bibliography{library,thomas,other}
\end{document}